\documentclass[prd, aps, nofootinbib, preprintnumbers, showpacs,
superscriptaddress,twocolumn,floatfix]{revtex4}

\usepackage{graphicx,epsfig,color}

\usepackage{amssymb,amsmath}
\usepackage[english]{babel}
\usepackage{graphicx}
\usepackage[latin9]{inputenc}
\usepackage{epstopdf}
\usepackage{color}

\newcommand{\pert}[2]{{}^{\mbox{\,\tiny $\{#1\}\!$}}{#2}}

\newcommand{\beq}{\begin{equation}}
\newcommand{\eeq}{\end{equation}}
\newcommand{\beqa}{\begin{eqnarray}}
\newcommand{\eeqa}{\end{eqnarray}}

\begin{document}

\title{Mode coupling of Schwarzschild perturbations: Ringdown frequencies}

\author{Enrique~Pazos}
\affiliation{Center for Relativistic Astrophysics, School of Physics, Georgia Institute of Technology, 837 State Street, Atlanta, GA 30332-0430 }
 \affiliation{Departamento de Matem\'atica, Universidad de San Carlos de Guatemala, Edificio T4, Facultad de Ingenier\'{\i}a, Ciudad Universitaria z. 12, Guatemala}

\author{David Brizuela}
\affiliation{Theoretisch-Physikalisches Institut,
Friedrich-Schiller-Universit\"at, Max-Wien-Platz 1, 07743 Jena, Germany}
  
\author{Jos\'e M. Mart\'{\i}n-Garc\'{\i}a}
\affiliation{Institut d'Astrophysique de Paris, CNRS, Univ. Pierre et Marie Curie,
98bis boulevard Arago, 75014 Paris, France}
\affiliation{
Laboratoire Univers et Th\'eories, CNRS, Univ. Paris Diderot,
5 place Jules Janssen, 92190 Meudon, France}

\author{Manuel~Tiglio}
\affiliation{Department of Physics, Center for Fundamental Physics, Center for Scientific Computation and Mathematical Modeling, Joint 
Space Institute. University of Maryland, College Park, MD 20742, USA.}
  
\begin{abstract}
Within linearized perturbation theory, black holes decay to their final stationary state through the well-known spectrum of quasinormal modes. Here we numerically study  whether nonlinearities change this picture. For that purpose we study the ringdown frequencies of gauge-invariant second-order gravitational perturbations induced by self-coupling of linearized perturbations of Schwarzschild black holes. We do so through high-accuracy simulations in the time domain of first and second-order Regge-Wheeler-Zerilli type equations, for a variety of initial data sets.  We consider first-order even-parity $(\ell=2,m=\pm 2)$ perturbations and odd-parity $(\ell=2,m=0)$ ones, and all the multipoles that they generate through self-coupling. For all of them and all the initial data sets considered we find that ---in contrast to previous predictions in the literature--- the numerical decay frequencies of second-order perturbations are the same ones of linearized theory, and we explain the observed behavior.  This would indicate, in particular, that when modeling or searching for ringdown gravitational waves, appropriately including the standard quasinormal modes already takes into account nonlinear effects. 
\end{abstract}


\maketitle

\section{Outlook and Motivation}
Black hole no-hair theorems~\cite{lrr-1998-6,Bekenstein:1998aw}
 state that within Einstein's theory the end point of any system with enough gravitational energy to form a black hole is remarkably simple: 
it is uniquely characterized by one member of the Kerr family\footnote{Charge is expected not
to play a significant role in most astrophysical scenarios.} \cite{Kerr63},
which is described by only two parameters: the spin and mass of the final
black hole.  

As a consequence, the details by which different systems decay to such endpoints have been of interest for many decades. 
Pioneering studies were done by a number of authors in the early seventies, starting with studies of linearized perturbations of non-rotating (Schwarzschild) black holes (e.g., \cite{Vishveshwara:1970zz,Vishveshwara70,Davis71}). Press realized  
that there is always an 
intermediate stage where the ringdown is dominated by a set of 
oscillating and exponentially decaying solutions, quasinormal modes (QNMs), whose spectrum depends only on the mass of the black hole and the multipole index $\ell$ of the initial perturbation \cite{Press:1971wr}. This regime is followed by a power-law 
`tail' decay due to backscattering \cite{Price72}. 

In the case of gravitational perturbations of non-rotating black holes the relevant equations from which QNMs can be inferred are the Regge-Wheeler \cite{Regge57}
and Zerilli \cite{Zerilli70,Zerilli70a} ones. For rotating black holes the corresponding one (though based on a curvature formalism, as opposed to a metric 
one) is the Teukolsky equation~\cite{Teukolsky73}. Their QNMs were first studied by Teukolsky and Press \cite{Press:1973zz}. 
See \cite{Berti:2009kk,Kokkotas99a_url} for comprehensive reviews on the rich area of QNMs.

The QNM with the lowest frequency is called the fundamental
one. Since the subsequent ones (overtones) decay much faster, the ringdown of Kerr black holes in linearized theory is in practice described by a few oscillating modes which decay exponentially 
in time, till they reach the tail regime. It is interesting to note that the tail decay problem for rotating black holes is still not completely understood  
 \cite{Krivan:1999wh,Poisson:2002jz,Gleiser:2007ti,Tiglio:2007jp,Burko:2007ju}. 

From an observational point of view this universal ringdown spectrum is of great power: one can use a single QNM detection to 
infer the mass and spin of the black hole source, assuming General Relativity to be correct. Alternatively, 
through a two-mode detection one can test General Relativity and/or the assumption that a black hole is the source of the 
measured signal \cite{Dreyer:2003bv}. The main idea is that the QNM frequencies of both detections have to be consistent 
with respect to their inferred masses/spins. 

The LISA mission is expected to measure gravitational waves in the low-frequency spectrum: $(10^{-5} - 10^{-1})$ Hz, such as 
those emitted in the collision of supermassive binary black holes 
(SMBBHs)~\cite{Berti:2005ys}. Flanagan 
and Hughes~\cite{Flanagan97a}  showed that, quite generically, the signal to noise ratio for these sources in the inspiral regime should 
be comparable to that one in the ringdown.  Therefore, detection of SMBBHs by LISA 
through the measurement of QNMs seems to be feasible. Assuming a lower cutoff of $(10^{-4}-10^{-5})$ Hz and 
requiring that the QNM signal lives long enough to travel once through 
LISA's propagation arms places a constraint on the mass range of the SMBBH candidates: a few $10^{5}M_\odot$  to $(10^8-10^9)M_{\odot}$. 

A step beyond detection analysis is that one of parameter estimation. 
In Ref.~\cite{Berti:2006cc,Berti:2005ys} it was found that through a single QNM detection LISA 
would be able to accurately infer the mass {\em and} spin of supermassive black holes: for black holes 
with mass $M\gtrsim 10^5M_{\odot}$ the errors in mass and spin would be smaller than one part in $10^2$, and smaller than one 
part in $10^5$  for the more optimistic case  $M\gtrsim 5 \times 10^5M_{\odot}$ \footnote{Only cases with  $M\gtrsim 10^5M_{\odot}$ are 
considered in   \cite{Berti:2005ys,Berti:2006cc,Berti:2005ys} because otherwise the QNM signal would be short lived enough that special detection 
techniques might be needed. }. These predicted accuracies depend on the ringdown efficiency $\epsilon_{rd}$, defined as the fraction of mass 
radiated in ringdown waves. In these references very conservative values were used: $\epsilon_{rd} \sim 0.1\%-3\%$. For example, 
 it has been found in numerical simulations of  two equal-mass, non-spinning black holes starting from quasi-circular motion that around $\epsilon_{rd} \sim 2\%-3\%$ of the total mass is radiated in the ringdown regime \cite{2007PhRvD..75l4018B}. The inclusion of different masses and/or spin increases this value 
 (see, for example, \cite{Berti:2007fi}). 

In \cite{Berti:2006cc,Berti:2005ys} it was also found that at least a second detection of 
{\em either} mass {\em or} spin should be possible for LISA. Resolving both spin and angular momentum
(or, equivalently, both frequency and damping times associated with the QNM oscillation of this second mode)
might require a very large critical signal to noise ratio,
which might in turn need the second mode to radiate a significant portion of the emitted gravitational wave when compared 
to the first one. Whether this is feasible or not can only be established by giving precise 
predictions of the amplitudes for secondary 
candidates. 

Underlying in all these analyses is the implicit assumption that quasinormal modes and their spectrum of associated frequencies accurately describe the (intermediate) stage of the ringdown to a final Kerr stationary state. Which is certainly the case in linearized theory, but at the same time there is evidence that effects of self-interaction in gravitational waves might be observable with the expected sensitivity of LISA \cite{Kocsis:PRD76}. 

Similarly, quasinormal modes also play an important role in the semi-analytical modeling of intermediate mass black holes (IMBH), such as in the Effective One Body approach \cite{BuonannoDamour:PRD62,BuonannoDamour:PRD59}, where the gravitational wave as modeled within this formalism is stitched to a ringdown one consisting of QNMs by enforcing continuity of the wave and its derivatives and using the values of quasinormal frequencies as expected from linearized theory \cite{Damour:2006tr,Buonanno:2006ui,Schnittman:2007ij}. In this context, it is worth recalling that in close-limit studies of binary black holes it was found that corrections from second-order perturbations were in some cases significant \cite{Gleiser96b}. For IMBHs, which could have total masses in the range of $\sim 100 M_{\odot} - 10^4 M_{\odot}$, it is especially important to accurately model the merger and ringdown since they should fall in the frequency band of earth based gravitational wave detectors. Although the existence of IMBHs is still debatable, they could provide an interesting source for Advanced LIGO and VIRGO if they are present in dense globular clusters \cite{Miller:2003sc} (see also \cite{:2010cfa}). Recent observational evidence of an IMBH can be found in \cite{Farrell:2009Nature,Farrell:2010jy,Webb:2010qg,Wiersema:2010ki}.

The previous discussions motivate us to study how nonlinear perturbations of black holes decay in time: do they do so just as in linearized theory or with a different spectrum of frequencies? We carry out our study through numerical simulations of first and second-order gauge-invariant gravitational perturbations of Schwarzschild black holes. We find that for all practical purposes, and to high numerical accuracy, the complex decay frequencies of second-order perturbations are the standard quasinormal ones from linearized theory --in contrast to previous predictions in the literature \cite{Ioka:2007ak,Nakano:2007cj,Okuzumi:2008ej}-- and we explain why this appears to be the case. Essentially, in all our simulations we find that mode-mode couplings excite nonlinearities in the early stages of the perturbations  {\em before} the quasinormal regime for the linearized perturbations kicks in. By the time the latter happens,  
those couplings have decreased to negligible values and the excited nonlinearities essentially propagate as in linearized theory; and, in particular, they decay with the standard QNM frequencies. 

The structure of the paper is as follows. Section~\ref{sec:formalism} reviews the basics of Regge-Wheeler-Zerilli equations and quasinormal modes, and Sec.~\ref{sec:formalism2} the main features of the gauge-invariant approach here used for second-order perturbations. Section~\ref{sec:numerics}
 describes our numerical approach and setup for solving the first and second-order Regge-Wheeler and Zerilli equations, and Sec.~\ref{sec:qnm} presents and discusses our main  results.
 
\section{First-order perturbations of Schwarzschild and QuasiNormal Modes} 
\label{sec:formalism}
Metric gravitational perturbations can be expanded in tensor spherical harmonics. At the linear level modes with different angular structure decouple from each other if the background spacetime is spherically symmetric, as is the case for the Schwarzschild metric. In addition, for each multipole, perturbations of Schwarzschild can be further split into two sectors with different parities, which also decouple from each other in linearized theory. 

For each multipole, each of these sectors is completely described by a master function which depends on time and radius. 
The Regge-Wheeler function
contains all the relevant information of the axial or odd-parity sector, whereas the Zerilli one encodes the polar or even degrees of freedom.
These functions satisfy master evolution equations, 
\begin{eqnarray}
\Box\pert{1}{\Phi}_\ell^m
-V_{\rm RW} \pert{1}{\Phi}_\ell^m &=& 0  \,, \label{eq:rw1} \\
\Box\pert{1}{\Psi}_\ell^m
-V_{\rm Z} \pert{1}{\Psi}_\ell^m &=& 0  \,, \label{eq:z1}
\end{eqnarray}
where $\pert{1}{\Psi}_\ell^m$ and $\pert{1}{\Phi}_\ell^m$ denote, 
respectively, the first-order Zerilli and Regge-Wheeler 
functions for a given $(\ell,m)$ mode.
The potentials that appear in these equations are
given by
\begin{equation}
V_{\rm Z}\equiv\frac{\ell(\ell+1)}{r^2}
-\frac{6M}{r^3}\frac{r^2\lambda(\lambda+2)+3M(r-M)}{(r\lambda+3M)^2},\,
\\
\end{equation}
\begin{equation}
V_{\rm RW}\equiv\frac{\ell(\ell+1)}{r^2}-\frac{6M}{r^3} 
\end{equation}
and are, in particular, independent of the azimuthal index $m$. In these expressions $\lambda\equiv\frac{1}{2}(\ell-1)(\ell+2)$, $M$ is the mass of the Schwarzschild black hole background, $r$ its 
areal radius, 
and $\Box$ is the two-dimensional D'Alambertian operator corresponding to the time and radial sector of the background.

The complete spectrum of QNMs can be numerically obtained by analyzing Eqs.~(\ref{eq:rw1},\ref{eq:z1}) in the 
frequency domain. Computing in this way the amplitudes of QNMs for any given initial data is not so straightforward, though. Bearing in mind 
our motivation of studying the behavior of second-order perturbations, we instead solve Eqs.~(\ref{eq:rw1},\ref{eq:z1}) in the time-domain. That is, we  prescribe (a variety of) initial data, evolve them and analyze the solutions at different observer locations as functions of time. 

\begin{figure}
\includegraphics[width=\columnwidth]{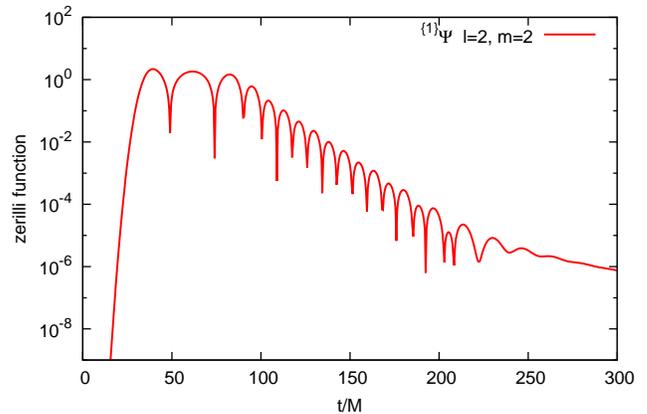}
\caption{Solution to the first-order Zerilli equation at observer location $r=51.8M$.}
\label{fig:solution-1st}
\end{figure}

The early behavior of the solution depends on the type of initial data, followed 
by the QNM ringdown of the black hole.
In Fig.~\ref{fig:solution-1st} we show a typical solution of the Zerilli equation,
for a fixed observer at $r=51.8M$ as a function of time. 
The initial data for this particular case consist of a Gaussian profile to the initial
time derivative of the Zerilli function, centered at $r=20M$ with
a width $\sigma=4M$, and the initial value of the Zerilli function itself is set to zero.
[This corresponds to what we call {\it time-derivative} initial data in the following sections, see Eqs.~(\ref{eq:td1},\ref{eq:td2})]. To measure the complex
QNM frequency we perform a numerical fit to a function of the
form
\begin{equation}
f(t)=A \, e^{at} \sin\left[b (t-t_0) \right] \, , 
\end{equation}
where the parameters to be fitted are $A$, $a$, $b$ and $t_0$ and the choice of the starting time for the QNM regime is chosen as that one which optimizes the fit, as introduced and explained in Ref.~\cite{Dorband06}. The quasinormal
frequency is therefore given as $\omega = a+bi$. The expected value for the fundamental QNM for an $\ell=2$
perturbation is $0.37367-0.08896i$~\cite{Kokkotas99a_url}, while our fit for this simulation yields $0.37077-0.08826i$. 

\section{Second-order gauge-invariant perturbations of Schwarzschild} 
\label{sec:formalism2}
We study second-order gravitational perturbations of Schwarzschild black holes using a gauge-invariant formalism for {\em arbitrary} first and second-order perturbations \cite{Brizuela:2009qd}. The key feature of this formalism is being able to consider perturbations with arbitrary angular multipole structure, and has been possible mostly due to the development of a
suitable theoretical framework \cite{Brizuela:2006ne, Brizuela:2007zza} and to the advance 
of very efficient symbolic algebra tools for tensor-type calculations  \cite{xAct,Brizuela:2008ra}. 

Next we very briefly summarize those results of the formalism presented in \cite{Brizuela:2009qd} which 
are relevant for the current work; see that reference for more details. 

Due to the intrinsic nonlinearities of General Relativity,
any non-trivial solutions of Eqs.~(\ref{eq:rw1},\ref{eq:z1}) generate second-order contributions which are solutions of  
 Zerilli and Regge-Wheeler-type equations with source terms,
\begin{eqnarray}
\Box\pert{2}{\Phi}_\ell^m-V_{\rm RW} \pert{2}{\Phi}_\ell^m=\pert{2}{\cal S}_\Phi \, , \label{eq:rw2}\\
\Box\pert{2}{\Psi}_\ell^m-V_{\rm Z} \pert{2}{\Psi}_\ell^m=\pert{2}{\cal S}_\Psi 
\label{eq:z2} \, .
\end{eqnarray}
The sources $\pert{2}{\cal S}_\Psi$ and $\pert{2}{\cal S}_\Phi$ depend quadratically on the lower-order 
perturbations and their time and space derivatives 
from both first-order sectors. That is, in general
the coupling of even (odd) parity modes generates
odd (even) parity second-order modes; see Appendix \ref{coupling}
for a detailed description of the selection rules for
this mode coupling. 

Second-order Regge-Wheeler-Zerilli (RWZ) type functions are not unique: one can add to them any quadratic combination of the first-order ones 
and they will still be gauge invariant and will still satisfy equations of the form (\ref{eq:rw2},\ref{eq:z2}) with different sources. For this reason, asking what are the ringdown frequencies of second-order RWZ functions is not an unambiguous question; as opposed to, say, asking what are the ringdown frequencies of the emitted gravitational power. It turns out, however, that, 
as discussed in \cite{Brizuela:2009qd} and made explicit below, since the second-order quantities, like $\pert{2}{\Psi}$, $\pert{2}{\Phi}$,
$\pert{2}{\cal S}_\Psi$ and $\pert{2}{\cal S}_\Phi$, that we use correspond 
to the ones labeled as {\it regularized} in Ref. \cite{Brizuela:2009qd}, there is a simple relationship between emitted power and the RWZ functions. Under such choice of second-order RWZ functions it is therefore unambiguous and physically motivated to ask what are their ringdown frequencies. 

Reference \cite{Brizuela:2009qd} deals with the most general case for these sources and computation of the radiated energy. Here 
we quantitatively explore the ringdown frequencies for some particular cases. 
Namely, we study first-order ($\ell=|m|=2$) even-parity and ($\ell=2, m=0$) odd-parity 
 perturbations and the modes that they generate through self-coupling.  
That is, for simplicity we do not consider the coupling between the mentioned
even and odd-parity first-order modes. We do not explicitly list the sources  $\pert{2}{\cal S}_\Psi$ and $\pert{2}{\cal S}_\Phi$ for the 
cases here considered because they are rather lengthy and complicated expressions, but they are available from the authors upon request. 
The generated modes and the radiated energy carried by them are described next and summarized in Table~\ref{tab:modes}. 
\begin{widetext}
\begin{center}
\begin{table}[h]
\begin{tabular}{cccc|cccc}
\hline
\hline
\multicolumn{2}{c}{First order} &&&&&  \multicolumn{2}{c}{Second order}   \\ 
Multipoles & Parity &&& && Multipoles & Parity \\ 
\hline
$(\ell=2 = |m|)$ & even &&&&&  $(\ell=4 =m)$,$(\ell=4, m=0)$,$(\ell=2, m=0)$ & even \\
$(\ell=2,m=0)$   & odd  &&&&&  $\phantom{(\ell=4 =m)}$ $(\ell=4, m=0)$,$(\ell=2, m=0)$ & even \\
\hline
\hline 
\end{tabular} \caption{First-order modes considered and second-order ones generated by self-coupling.}\label{tab:modes}
\label{tab:modes2}
\end{table}
\end{center}
\end{widetext}

\subsection{CASE A: even-parity $\ell=|m|=2$ perturbations and generated modes}  \label{sec:casea}
As discussed in Appendix \ref{coupling},
the self-coupling between these modes generates second-order
$(\ell=4, m=\pm 4)$ even-parity (polar) ones, whereas the coupling
between them ($m=2$ with $m=-2$) gives rise to second-order $(\ell=4,m=0)$,
$(\ell=2,m=0)$ and $(\ell=0,m=0)$ even-parity (polar) modes, and  
 $(\ell=3,m=0)$ and $(\ell=1,m=0)$ odd-parity (axial) ones.
Since this paper only deals with different radiative aspects of the 
system, we can ignore modes with $\ell<2$. 

Furthermore, we assume that the Zerilli functions
$\pert{1}{\Psi}_2^{\pm 2}$ that
describe the first-order $(l=2,m=\pm 2)$ modes are real
\begin{equation}
\pert{1}{\Psi}_2^{\pm 2} \in \mathbb{R} \, .  \label{eq:real}
\end{equation}
This means that both modes are described by the same function, since
generically the relation $(\pert{1}{\Psi}_l^m)^*=(-1)^m\pert{1}{\Psi}_l^{-m}$
holds, and, in essence,
the system reduces to a problem of self-coupling. The second-order
even-parity (polar) modes inherit this property, in such a 
way that we only need three functions to describe them:
$$
\{\pert{2}{\Psi}_2^0, \pert{2}{\Psi}_4^0, \pert{2}{\Psi}_4^{4}\} \, .
$$
Due to the assumption (\ref{eq:real}), 
none of the second-order odd-parity (axial) modes are generated.
This happens because the source for a $m=0$ mode must be real.
Schematically, the generic term of the source for
this axial $(l=3,m=0)$ mode can be written as
$i\pert{1}{\Psi}_2^2\pert{1}{\Psi}_{2}^{-2}$
and it is straightforward to see that its real part vanishes
under the assumption (\ref{eq:real}).

In this particular CASE A, the radiated power associated with the mentioned
modes for a given observer located at $r_{obs}$ as a function of time is given by
\begin{widetext}
\begin{equation}
{\rm Power} (r_{obs},t) =  \frac{dE}{dt} = 
\frac{\epsilon^2}{12\pi}\left|\partial_t\pert{1}{\Psi}_2^2\right|^2
+\frac{9\epsilon^4}{640\pi}\left\{ 2\left|\partial_t\pert{2}{\Psi}_4^4\right|^2 + \left|\partial_t\pert{2}{\Psi}_4^0\right|^2
\right\}
+\frac{\epsilon^4}{96\pi}\left|\partial_t\pert{2}{\Psi}_2^0\right|^2 + {\cal O} \left( \epsilon ^5 \right) \ , 
\label{eq:power_even}
\end{equation}
\end{widetext}
where $\epsilon$ is the perturbative parameter and all the expressions on the right-hand side are evaluated, naturally, at $(r_{obs},t)$.
In principle this equation is valid only at null infinity but, as it is usually the case in computations, we evaluate it 
at a finite radius.

\subsection{CASE B: odd-parity ($\ell=2,m=0$) perturbations and generated modes}  \label{sec:caseb}

In principle, following the selection rules,
the odd parity ($\ell=2,m=0$) mode would generate by self-coupling the
second-order ($\ell=0,m=0$), ($\ell=2,m=0$) and ($\ell=4,m=0$) even-parity modes, as
well as the ($\ell=1,m=0$) and ($\ell=3,m=0$) odd-parity ones. However, in this
axisymmetric case ($m=0$ for all modes), the Clebsch-Gordan-like coefficients
that appear in the sources $\pert{2}{\cal S}_\Phi$ and $\pert{2}{\cal S}_\Psi$
vanish for those second-order modes with an odd harmonic label $\ell$
(see Appendix \ref{coupling}). Hence, none of the second-order odd-parity
modes are excited and, up to this order, we are left with
two radiative second-order modes, 
$$
\{\pert{2}{\Psi}_4^0, \pert{2}{\Psi}_2^0 \} \, .
$$
The radiated power is then given by 
\begin{widetext}
\begin{equation}
{\rm Power} (r_{obs},t) =
\frac{3\epsilon^2}{32\pi}\left|\partial_t\pert{1}{\Phi}_2^0\right|^2 + 
\frac{3\epsilon^4}{32\pi}\left\{
\frac{3}{20}\left|\partial_t\pert{2}{\Psi}_4^0\right|^2 + 
\frac{1}{9}\left|\partial_t\pert{2}{\Psi}_2^0\right|^2
\right\} 
+{\cal O}(\epsilon^5). \label{eq:power_odd}
\end{equation}
\end{widetext}

\section{Numerical approach for solving the master equations} \label{sec:numerics}

We now describe in some detail our numerical approach for evolving the first and second-order RWZ equations, since in the 
past difficulties have been reported with the high-order derivatives in the sources of these equations. 

We numerically solve the first and second-order equations using a pseudo-spectral collocation method.
The spatial derivatives are computed using Chebyshev polynomials and Gauss-Lobatto (GL) collocation points, and 
the system is evolved in time using a standard fourth-order Runge-Kutta scheme. We use a small enough timestep for the 
time integration so that the solution converges exponentially with the number of collocation points (see below).  
 The accuracy of all the simulations presented in this paper are at the level of double precision round-off. 
 
GL collocation points are not equally spaced; instead, they cluster near the edges of the computational domain 
(equally spaced points would not give exponential convergence). For that reason it is standard to use a multi-domain 
approach. Here we subdivide our radial domain in (non-overlapping) blocks of 
length $10M$ each, communicated through a penalty technique. At the interface each 
incoming characteristic mode $u^{+}$ is penalized according to (see \cite{hesthaven3} and references therein) 
$$
\dot{u^{+}} = \left( \ldots \right) - \frac{\alpha N^2\delta}{r_{block}}(u^{+}-v^{+})
$$
where $v^{+}$ is the value of the same mode at the interface point using the
neighboring block, 
$r_{block}$ is the size of the corresponding block ($10M$ in these
simulations), $\alpha $ is the associated characteristic speed, $N$ the
number of collocation points on that block and $\delta$ a penalty parameter
chosen here to be $\delta = 0.6$. At the outer boundary each characteristic incoming mode is
similarly penalized to zero; though this is done simply to achieve stability, in our simulations the domain is large 
enough that our results are causally disconnected from the outer boundary. The singularity of the black hole is 
dealt with through excision. That is, by using Kerr-Schild coordinates for the background spacetime and placing an inner boundary inside the event horizon.

\begin{figure}
\includegraphics[width=\columnwidth]{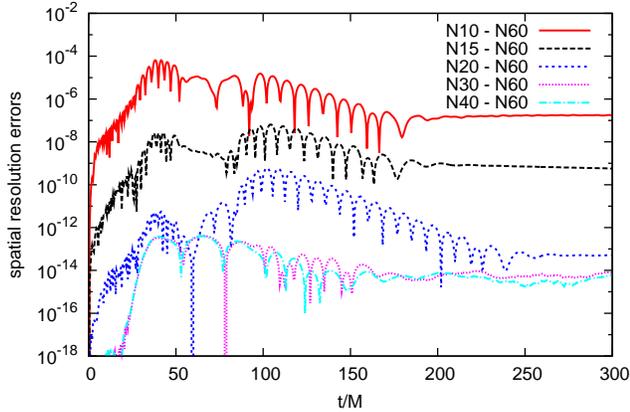}
\includegraphics[width=\columnwidth]{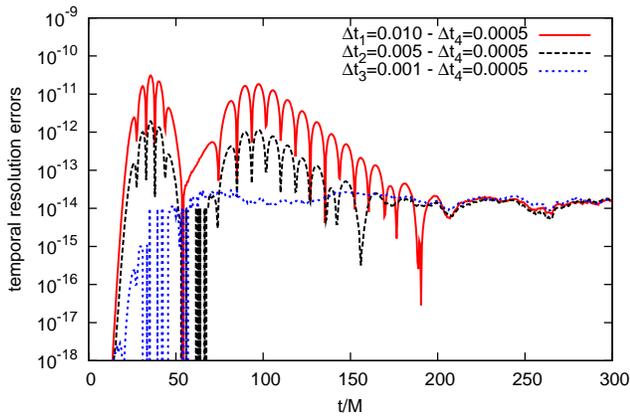}
\caption{Numerical errors for different spatial resolutions 
using a fixed timestep $\Delta t=0.01M$ (top),
and for different timesteps using a fixed spatial
resolution of $N=60$ points per domain (bottom). Both figures 
show the differences between 
several resolutions and the most accurate one, which is
$N=60$ for the top panel and $\Delta t_4=0.0005M$ for the
bottom one. In both cases the observer is located at $r=51.8M$.
We see exponential convergence and errors in the order of
double precision round-off.} 
\label{fig:errors-dtdr}
\end{figure}

\begin{figure}
\includegraphics[width=\columnwidth]{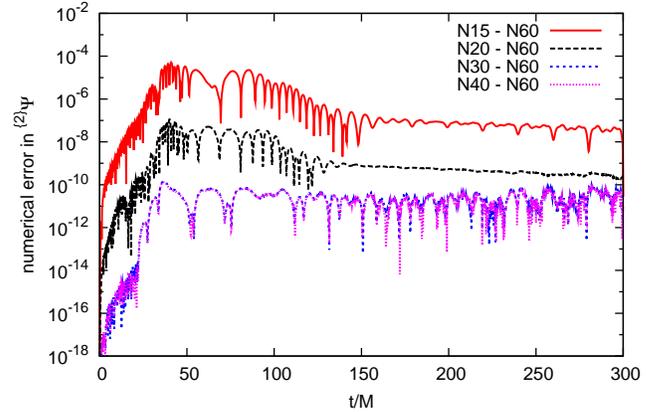}
\caption{Numerical errors in the second-order Zerilli function 
$\pert{2}\Psi_4^4$ for different spatial resolutions and
a fixed timestep $\Delta t=0.001M$.
The errors are to be interpreted as in the previous figures. For $N=30$ (which are the ones used in the rest of this paper) and higher,  
they are at the level of double precision round-off. } 
\label{fig:errors-dtdr2}
\end{figure}

As an example, Fig.~\ref{fig:errors-dtdr} shows a self-convergence test for the first-order $(\ell =2 = m)$ Zerilli function, extracted at 
 $r=51.8M$ \footnote{We place observers at the beginning/end of each domain: $1.8M,11.8M,21.8M$, etc. }, both changing the number 
of collocation points as well as the timestep. 
The initial data used for such test correspond to the same one used in Fig.~\ref{fig:solution-1st},    
\begin{equation}
\pert{1}{\Psi}_2^2=0 \;\;\; , \;\;\;\ \pert{1}{\dot{\Psi}}_2^2(t=0,r) = e^{-(r-r_0)^2/\sigma^2}\, , 
\end{equation}
with $\sigma = 4M$, $r_0=20M$, and the spatial domain $r \in[1.8M,301.8M]$. In Fig.~\ref{fig:errors-dtdr2} we also show the result of a 
 convergence test for the generated 
 second order $(\ell = 4 = m)$ Zerilli mode. From these figures we see 
that using $30$ collocation points per domain and a timestep $\Delta t = 0.001M$ gives a numerical error at the level 
of double precision round-off; from hereon we use such resolutions for all our simulations. 

\begin{figure}
\includegraphics[width=\columnwidth]{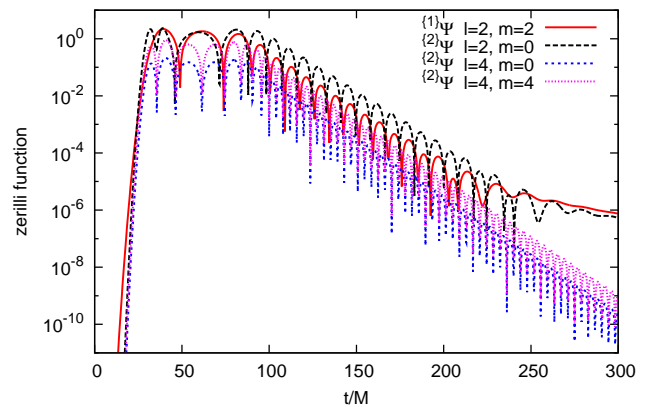}
\caption{Absolute value of first and second-order Zerilli functions extracted at $r=51.8M$.}
\label{fig:allZ}
\end{figure}

In order to compare the magnitude of the errors with the solutions themselves, in Fig.~\ref{fig:allZ} we show the absolute values 
of the first-order $\pert{1}\Psi_2^2$ and second-order $\{ \pert{2}\Psi_2^0, \pert{2}\Psi_4^0 , \pert{2}\Psi_4^4 \}$ Zerilli solutions from the previous plots at their highest resolutions, all extracted at the same observer location. 

We note in passing that for most of the ringdown 
the order of magnitude of the second-order Zerilli functions appears to be comparable to
(and in one case even larger than) the first-order one. There is no contradiction in this, since their contribution to the radiated energy is scaled by $\epsilon ^4$,  while the contribution of the first order Zerilli function is scaled by $\epsilon ^2$, see Eq.~(\ref{eq:power_even}). 

\subsection{Setup of numerical simulations} 
\label{sec:setup}
We could introduce non-vanishing second-order modes via initial second-order perturbations. However,
we are interested in mode-mode coupling. Put differently, we are interested in the particular solution of Eqs.~(\ref{eq:z2},\ref{eq:rw2}) (that is, the one with vanishing initial data), since
the homogeneous one will be exactly the same as at first order.
Therefore, in this paper we always impose 
vanishing initial data for all the second-order modes and concentrate on those modes
generated by first-order mode coupling.

In the following section we solve the first-order RWZ equations with four different types of initial data, varying both the location $r_0$ and width 
$\sigma$ of the initial data, 
\begin{enumerate}
\item { Time Derivative (TD)}
\begin{eqnarray}
\pert{1}{\Psi}_2^2(t=0,r) &=&0 \, , \label{eq:td1} \\
\pert{1}{\dot{\Psi}}_2^2(t=0,r) &=& e^{-(r-r_0)^2/\sigma^2}. \label{eq:td2}
\end{eqnarray}
\item { Time Symmetric (TS)}
\begin{eqnarray*}
\pert{1}{\Psi}_2^2(t=0,r) &=& M e^{-(r-r_0)^2/\sigma^2} ,\\\
\pert{1}{\dot{\Psi}}_2^2(t=0,r) &=& 0 \,.
\end{eqnarray*}
\item { Approximately Outgoing (OUT)}
\begin{eqnarray*}
\pert{1}{\Psi}_2^2(t=0,r) &=& M e^{-(r-r_0)^2/\sigma^2} ,\\
\pert{1}{\dot{\Psi}}_2^2(t=0,r)  &=& - (1-2M/r) \partial_r \pert{1}{\Psi_2^2}(t=0,r) .
\end{eqnarray*}
\item {Approximately Ingoing (IN)}
\begin{eqnarray*}
\pert{1}{\Psi}_2^2(t=0,r) &=& M e^{-(r-r_0)^2/\sigma^2} ,\\
\pert{1}{\dot{\Psi}}_2^2(t=0,r)  &=& (1-2M/r) \partial_r \pert{1}{\Psi_2^2}(t=0,r) .
\end{eqnarray*}
\end{enumerate}

\section{Decay frequencies of second-order perturbations}
\label{sec:qnm}

Ioka and Nakano have put forward the suggestion that at second order
new frequencies should appear in the ringdown spectrum, which
 would be given by the sum of different pairs of standard QNM frequencies. In particular, according to this prediction, the dominant frequency would correspond to the 
double of the standard fundamental one from linearized theory \cite{Ioka:2007ak,Nakano:2007cj}. 
This seems reasonable, since the sources for the second-order master equations are quadratic in the
first-order modes, so one might well expect that frequencies get summed-up in Fourier space. The physical 
picture, however, appears to be at the same time more subtle and simpler: our numerical simulations indicate that 
in practice second-order perturbations decay with the standard QNM frequencies from linearized theory. 

Recall that the physical process here studied is the coupling of linear modes. That is, 
 we initialize all second-order perturbations to zero. 
The second-order master equations have sources which are, indeed, quadratic in the first-order modes. What we observe 
in all our simulations, though,  is that those sources quickly excite the second-order perturbations and afterwards  
decay very fast in time. As a consequence, once the second-order modes have been excited and reached the regime in which they 
oscillate with a constant complex frequency (i.e. what in linearized theory corresponds to the QNM regime), they essentially propagate 
with a vanishing source. In other words, as first-order perturbations do. And, in particular, they 
do oscillate and decay with the same, standard, QNM frequencies from linearized perturbation theory. 

\begin{figure}
\includegraphics[width=0.9\columnwidth]{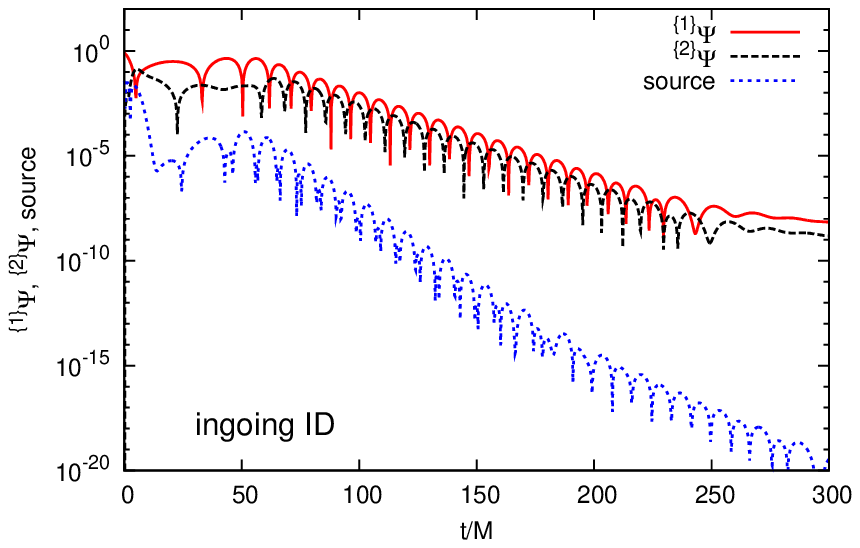}
\includegraphics[width=0.9\columnwidth]{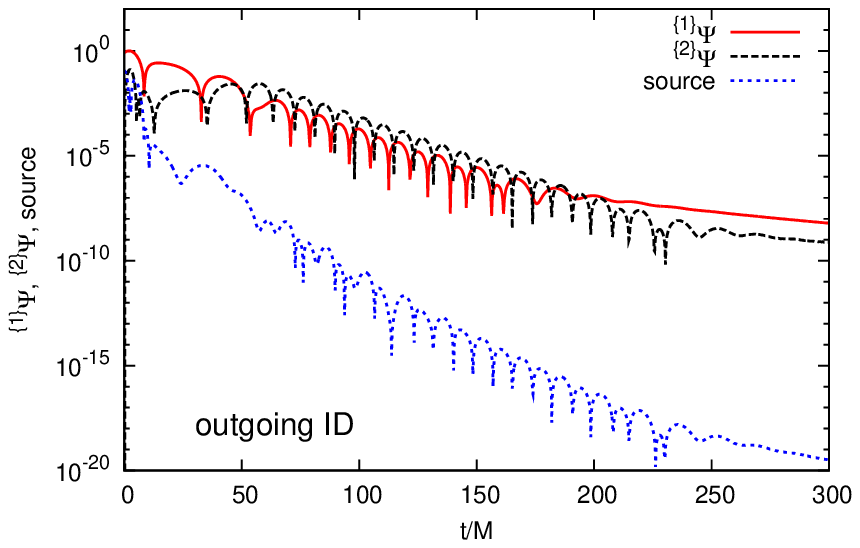}
\includegraphics[width=0.9\columnwidth]{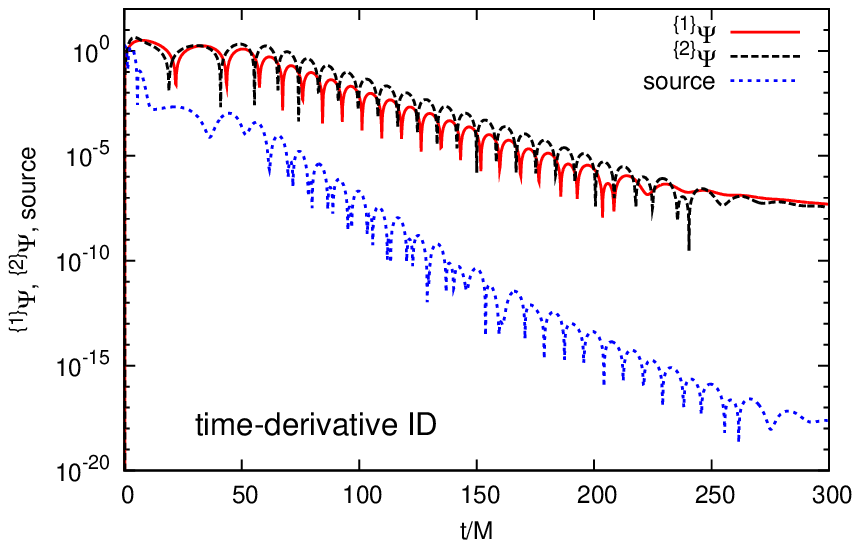}
\includegraphics[width=0.9\columnwidth]{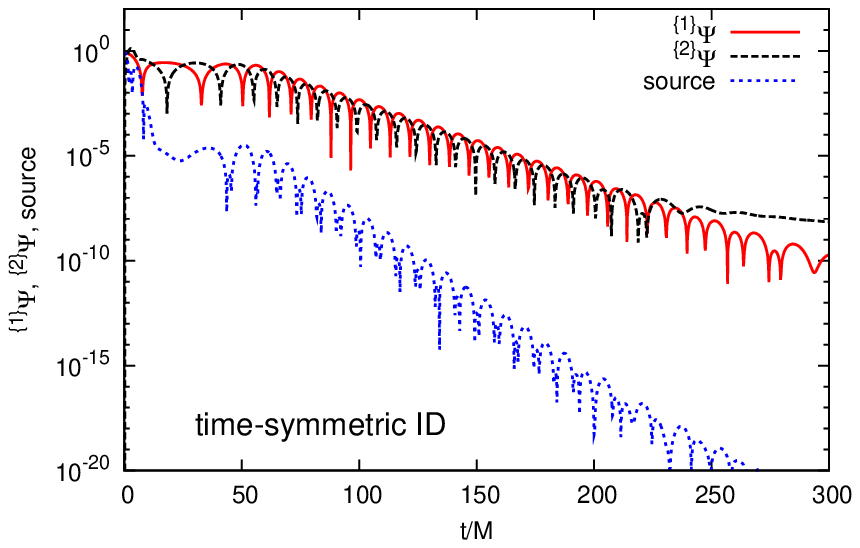}
\caption{CASE A simulations:  first-order ($\ell=2, m=2$) Zerilli function 
and second-order ($\ell=2, m=0$) one, along
with the source term for the second-order master equation [Eq.~(\ref{eq:z2})], as functions of time for different initial data profiles. The source plays a role 
only at very early times in exciting the second-order modes, afterwards being much smaller than the first and second-
order solutions.}
\label{fig:z1z2-source}
\end{figure}

We show this behavior in some detail for the four initial data types (Sec.~\ref{sec:setup}) of CASE A perturbations (Sec.~\ref{sec:casea}), with $r_0=20M$ and $\sigma=4M$. 
Figure \ref{fig:z1z2-source} shows the first-order Zerilli function and the ($\ell=2,m=0$) second-order one  for different initial data profiles. In all  
cases the source decays much faster than the second-order solution itself and
  therefore its role in determining the behavior of the latter by the time it enters the QNM regime is negligible. 
From the same figure one can notice that the type of initial perturbation does determine the time by which the second-order Zerilli function enters the tail regime; but this is not surprising, since it already happens for the first-order one. 

\begin{figure}
\includegraphics[width=0.9\columnwidth]{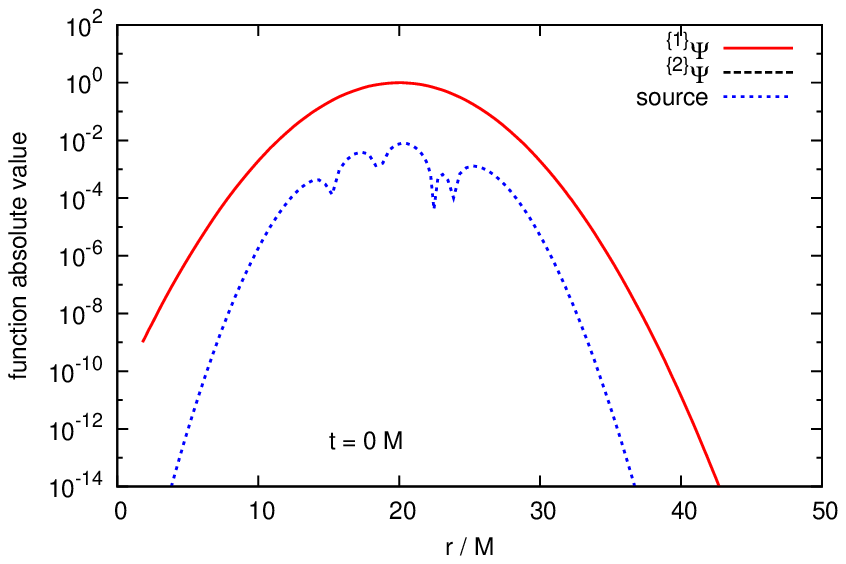}
\includegraphics[width=0.9\columnwidth]{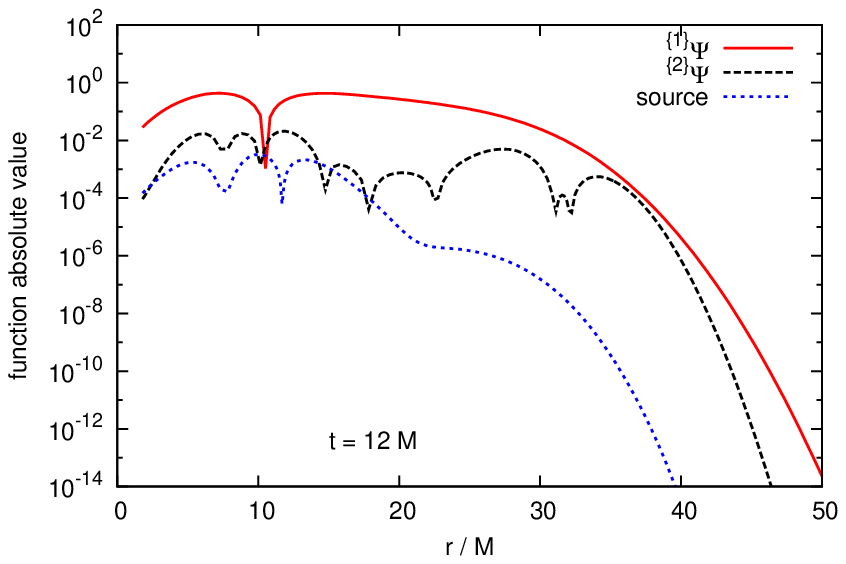}
\includegraphics[width=0.9\columnwidth]{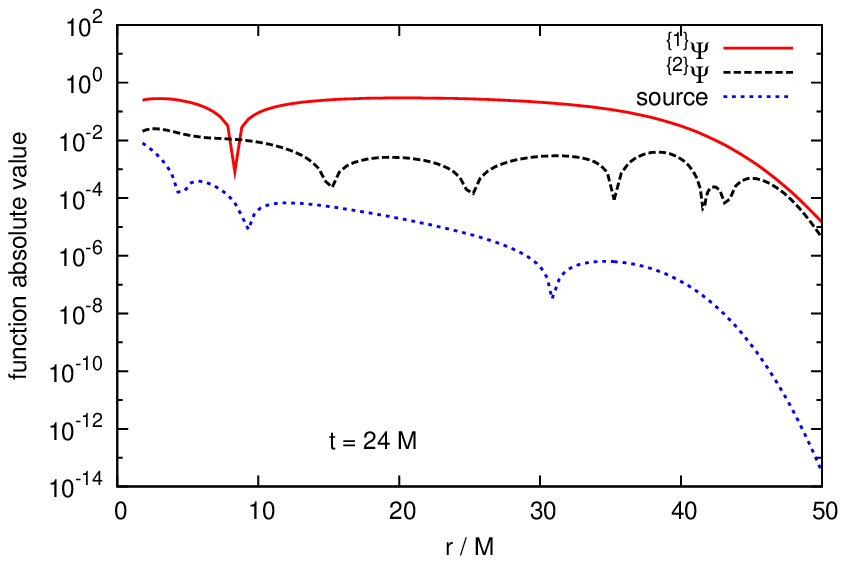}
\caption{CASE A simulations: snapshots of the first-order Zerilli function 
and second order ($\ell=4, m=0$) one, along 
with the source term, for ingoing initial data. The generic behavior of the source is to
rapidly decrease several orders of magnitude below the solutions 
themselves. [Note: In the first snapshot the second-order Zerilli function vanishes because it corresponds to $t=0$ and our 
setup of the physical problem.]}
\label{fig:snapshots}
\end{figure}

In order to gain further insight into these observations we display the dynamics of
first and second-order Zerilli functions and the source term $\pert{2}{\cal{S}}_\Psi$ of 
Eq.~(\ref{eq:z2}), now for the ($\ell=4,m=0$) second-order mode. Fig.~\ref{fig:snapshots} shows these three quantities
as functions of radius at different times. The source term is dominant only during the
first $\sim 20M$, later decaying at a fast rate to several orders of magnitude
below the second-order Zerilli function.

Table \ref{tab:qnm} shows the fitted QNM frequencies from our numerical data, for the different initial-data profiles of CASE A perturbations, with $r_0=20M$ and $\sigma=4M$. They 
agree quite well with those predicted by first-order theory for each of those modes. 

\begin{figure}
\includegraphics[width=0.9\columnwidth]{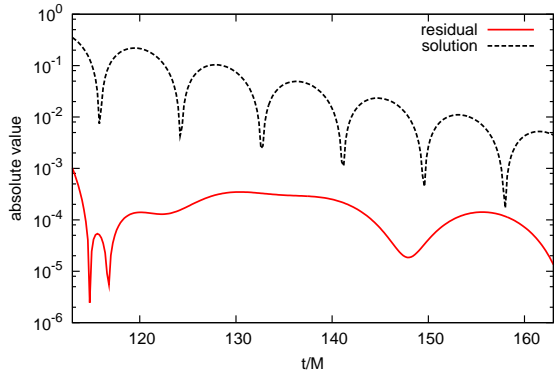}
\caption{Residual from fitting a second-order Zerilli function to a complex frequency mode. The fitted frequency corresponds to the standard fundamental quasinormal one for that multipole index, and the residual does not appear to contain traces of twice that frequency (or any other). }
\label{fig:residual}
\end{figure}

\begin{table}
\begin{tabular}{lcc|cc}
\hline
\hline
   & \multicolumn{2}{c}{1st order}  & \multicolumn{2}{c}{2nd order} \\
ID & $\ell=2$, $ m=2$ &err \% & $\ell=2$, $m=0$ &err \%  \\ 
\hline
TD & $0.37077-0.08826i$ & 0.8 &$0.37334-0.08883i$ & 0.1 \\ 
TS & $0.37353-0.08837i$ & 0.2 &$0.37335-0.08766i$ & 0.3 \\ 
IN & $0.37061-0.08887i$ & 0.8 &$0.37373-0.08945i$ & 0.1 \\ 
OUT& $0.37107-0.08624i$ & 1.0 &$0.37074-0.08902i$ & 0.8 \\ 
\hline
\hline
\\
\\
\end{tabular}

\begin{tabular}{lcc|cc}
\hline
\hline
   &  \multicolumn{4}{c}{2nd order}  \\
ID & $\ell=4$, $m=0$ &err \% & $\ell=4$, $m=4$ &err \% \\
\hline
TD & $0.80916-0.09418i$ & 0.003 & $0.80916-0.09418i$ & 0.003 \\
TS & $0.80920-0.09420i$ & 0.005 & $0.80920-0.09420i$ & 0.005 \\
IN & $0.80918-0.09416i$ & 0     & $0.80918-0.09416i$ & 0 \\
OUT& $0.80931-0.09425i$ & 0.019 & $0.80931-0.09425i$ & 0.019\\
\hline
\hline
\end{tabular}
\caption{Measured quasinormal frequencies from our numerical simulations (CASE A). 
They agree with those predicted by 
linearized theory, {\em even} for the second-order modes generated due to mode-mode coupling. 
The predicted QNM frequencies from standard linearized perturbation 
theory, as quoted in ~\cite{Kokkotas99a_url}, are $0.37367-0.08896i$ for $\ell=2$ and
$0.80918-0.09416i$ for $\ell=4$ (in linearized theory there is degeneracy with respect to the azimuthal index $m$. The relative errors for real and
imaginary parts of the measured frequencies $\omega$ are computed as 
$|\omega - \omega_{\mbox{exact}}|/|\omega_{\mbox{exact}}|$.}
\label{tab:qnm}
\end{table}
As described in Sec.~\ref{sec:numerics}, our numerical simulations are of very high accuracy (both at first and second-order): all the errors 
are at the level of double precision round-off, $\sim 10^{-14}-10^{-12}$ (see Figs.~\ref{fig:errors-dtdr},\ref{fig:errors-dtdr2}). 
Therefore, we do not consider lack of resolution as a possible reason for not finding   
traces of the predicted new second-order QNM frequencies in our simulations. Similarly, one might think that those predicted frequencies are in fact present, but with a  very 
small amplitude. Figure \ref{fig:residual} indicates that in practice that does not seem to be the case: the residual of the fit for the second-order Zerilli function [in the case shown it is the $(\ell=2,m=0)$ one, for TD initial data, first-order perturbations] ---defined as the function minus its fit--- does not appear at all to correspond to an oscillation and decay with twice the standard complex fundamental quasinormal frequency for that mode (or to an overtone). Instead, it appears to be the residual associated with the fact that QNMs are not complete.  

For completeness, in Appendix \ref{sec:qnm2} we provide results of the fitted frequencies for the four initial data types of CASE A perturbations, now varying both the location and width of the initial data; all of them support the same conclusion. 

Finally, we briefly discuss the results of some CASE B [odd-parity $(\ell=2,m=0)$ first-order mode] perturbations, since the conclusions are identical. As discussed in Sec.~\ref{sec:caseb} and summarized in Table~\ref{tab:modes2}, they generate both $(\ell=4,m=0)$ and $(\ell=2, m=0)$ even-parity second-order modes. The fitted frequencies from the numerical solutions (for a simulation of TD linear initial data with $r_0=20M$ and $\sigma=4M$) for these second-order modes yield, respectively, $0.37441-0.08921i$ and $0.80932-0.09419i$, to be compared with the expected values from perturbation theory: $0.37367-0.08896i$ for $\ell=2$ and $0.80918-0.09416i$ for $\ell=4$.

\section{Final remarks}
In this paper we have numerically evolved first and second-order self-generated gauge-invariant gravitational perturbations of Schwarzschild black holes with a variety of initial data sets, studying the oscillation and decay behavior of nonlinear modes and, more specifically, whether they correspond to the standard QNM frequencies or to a different spectrum. We have found, in all cases, that second-order modes decay through the standard QNM frequencies, and that the picture behind this is remarkably simple: first-order perturbations trigger high-order ones through source terms which afterwards rapidly decay in time. Besides, by the time the solutions reach the regime in which they oscillate and decay at a constant rate (the QNM regime in the case of linearized perturbations), the second-order modes for all practical purposes propagate as in linearized theory. 

Mode-mode coupling in the ringdown of black holes has been previously studied through numerical simulations of full Einstein equations \cite{Allen:1998wy,Zlochower03}; however, no conclusions seem to have reached or sought for in terms of the deviations in the ringdown spectrum from the linearized one (presumably due to lack of resolution). 

The fact that nonlinear aspects of Einstein's equations in the ringdown of black holes appear to already be captured ---at least in what oscillation and decay frequencies concerns--- by linearized perturbation theory is somehow remarkable and should be of use both for modeling black holes in the ringdown regime as well as in data analysis searches of gravitational waves. 

\begin{acknowledgments}

This research has been supported in part by NSF Grant PHYS 0801213 to the University of Maryland, NSF PHYS 0925345, 0941417, 0903973, 0955825
to Georgia Tech, the Spanish MICINN Project FIS2008-06078-C03-03,
the French A.N.R. Grant {\it LISA Science} BLAN07-1\_201699,
and by the Deutsche Forschungsgemeinschaft (DFG)
through SFB/TR 7 ``Gravitational Wave Astronomy''.

We thank Emanuele Berti, Vitor Cardoso, Chad Galley and Bence Kocsis for helpful discussions and suggestions. 
We also thank Carlos Ann for inspiration, Tryst DC --where parts of this work were done-- for hospitality, and Eric Cartman for proof-reading the manuscript and providing critical comments. 
\end{acknowledgments}

\appendix

\section{Mode coupling: selection rules}\label{coupling}

In this appendix we summarize, for completeness,
the selection rules for mode coupling \cite{Brizuela:2006ne, Brizuela:2007zza}.
A second-order $(l,m)$-mode gets a contribution from
a pair of first-order modes $(\hat l, \hat m)$ and $(\bar l, \bar m)$
if three conditions are obeyed. First, the
harmonic labels must be related by the usual
composition formulas
\begin{equation}\label{composition}
|\hat l-\bar l|\leq l \leq \hat l+\bar l,\quad{\rm and}\quad
\hat m + \bar m = m.
\end{equation}
Second, mode coupling must conserve parity.
To any harmonic coefficient with label $l$, we
associate a polarity sign $\sigma$ such that, under
parity, the harmonic changes by a sign $\sigma(-1)^l$.
Polar/even parity (axial/odd parity) harmonics have $\sigma=+1$
($\sigma=-1$). Then, parity conservation implies the
third condition:
\begin{equation} \label{parity}
(-1)^{\bar l+\hat l-l}
=\sigma\bar\sigma\hat\sigma,
\end{equation}
where $\hat\sigma$ and $\bar\sigma$ are the polarity
signs corresponding to the modes $(\hat l, \hat m)$ and $(\bar l, \bar m)$
respectively.
In particular, there is a special case in which the
coupling of two modes satisfying Eqs.
(\ref{composition}) and (\ref{parity}) does not
contribute to a second-order mode, and the reason
comes from the properties of the Clebsch-Gordan-like
coefficients that appear in the product formula for
the tensor harmonics \cite{Brizuela:2006ne}. In axisymmetry
($\bar m=\hat m=0$) the mentioned Clebsch-Gordan-like coefficients 
vanish if $\bar l+\hat l+l$ is odd.

This analysis can be extended to higher orders. In
particular, the parity condition implies that a
collection of $k$ modes with harmonic labels
$\{l_1,...,l_k\}$ and polarities
$\{\sigma_1,...,\sigma_k\}$ contributes to the
mode $(l,\sigma)$ only if $(-1)^l\sigma =
\Pi_{i=1}^{k}(-1)^{l_i}\sigma_i$.

\begin{widetext}
\section{Numerical decay frequencies of second-order perturbations from time domain simulations}\label{sec:qnm2}
\begin{center}
\begin{table}[h]
\begin{tabular}{rccccccc|ccccc}
\hline \hline
        && \multicolumn{5}{c}{Ingoing}                                && \multicolumn{5}{c}{Outgoing} \\
$r_0$   && $\ell=2$, $m=0$   && $\ell=4$, $m=0$ && $\ell=4$, $m=4$    && $\ell=2$, $m=0$ && $\ell=4$, $m=0$ && $\ell=4$, $m=4$ \\
 20     && $0.37373-0.08949i$ && $0.80918-0.09416i$ && $0.80918-0.09416i$ && $0.37074-0.08902i$ && $0.80931-0.09425i$ && $0.80931-0.09425i$ \\
 40     && $0.37147-0.08833i$ && $0.80916-0.09417i$ && $0.80916-0.09417i$ && $0.37079-0.08891i$ && $0.80911-0.09409i$ && $0.80911-0.09409i$ \\
 60     && $0.37141-0.08834i$ && $0.80915-0.09413i$ && $0.80915-0.09413i$ && $0.37093-0.08844i$ && $0.80911-0.09417i$ && $0.80911-0.09417i$ \\
 80     && $0.37102-0.08839i$ && $0.80917-0.09415i$ && $0.80917-0.09415i$ && $0.37140-0.08816i$ && $0.80913-0.09413i$ && $0.80913-0.09413i$ \\
100     && $0.37101-0.08840i$ && $0.80888-0.09414i$ && $0.80872-0.09417i$ && $0.37148-0.08782i$ && $0.80914-0.09426i$ && $0.80914-0.09426i$ \\
\hline
        &&                 &&                 &&                 &&                 &&                 &&    \\
\hline
$\sigma$&&                 &&                 &&                 &&                 &&                 &&    \\
 2      && $0.37708-0.09116i$ && $0.80918-0.09418i$ && $0.80918-0.09418i$ && $0.37365-0.08891i$ && $0.80918-0.09416i$ && $0.80918-0.09416i$ \\
 4      && $0.37373-0.08949i$ && $0.80919-0.09417i$ && $0.80918-0.09417i$ && $0.37079-0.08902i$ && $0.80931-0.09425i$ && $0.80911-0.09409i$ \\
 8      && $0.37286-0.08893i$ && $0.80875-0.09407i$ && $0.80875-0.09407i$ && $0.36887-0.08284i$ && $0.80923-0.09599i$ && $0.80921-0.09599i$ \\
\hline \hline
      && \multicolumn{5}{c}{Time Derivative}                                && \multicolumn{5}{c}{Time symmetric} \\
$r_0$ && $\ell=2$, $m=0$   && $\ell=4$, $m=0$ && $\ell=4$, $m=4$ && $\ell=2$, $m=0$ && $\ell=4$, $m=0$ && $\ell=4$, $m=4$ \\
 20     && $0.37334-0.08883i$ && $0.80916-0.09418i$ && $0.80916-0.09418i$ && $0.37335-0.08766i$ && $0.80920-0.09420i$ && $0.80920-0.09420i$ \\
 40     && $0.37277-0.08893i$ && $0.80919-0.09417i$ && $0.80919-0.09417i$ && $0.37376-0.08984i$ && $0.80915-0.09413i$ && $0.80915-0.09413i$ \\
 60     && $0.37378-0.08921i$ && $0.80920-0.09418i$ && $0.80920-0.09418i$ && $0.37267-0.08948i$ && $0.80917-0.09416i$ && $0.80917-0.09416i$ \\
 80     && $0.37383-0.08995i$ && $0.80915-0.09411i$ && $0.80915-0.09411i$ && $0.37344-0.08852i$ && $0.80917-0.09413i$ && $0.80917-0.09413i$ \\
100     && $0.37387-0.08999i$ && $0.80914-0.09409i$ && $0.80914-0.09409i$ && $0.37321-0.08823i$ && $0.80923-0.09422i$ && $0.80923-0.09422i$ \\
\hline
        &&                 &&                 &&                 &&                 &&                 &&    \\
\hline
$\sigma$&&                 &&                 &&                 &&                 &&                 &&    \\
 2      && $0.37360-0.08906i$ && $0.80917-0.09417i$ && $0.80917-0.09417i$ && $0.37404-0.08972i$ && $0.80917-0.09416i$ && $0.80916-0.09416i$ \\
 4      && $0.37334-0.08893i$ && $0.80916-0.09417i$ && $0.80916-0.09418i$ && $0.37335-0.08766i$ && $0.80920-0.09420i$ && $0.80920-0.09420i$ \\
 8      && $0.37315-0.08935i$ && $0.80924-0.09342i$ && $0.80924-0.09334i$ && $0.37943-0.08760i$ && $0.80867-0.09424i$ && $0.80867-0.09423i$ \\
\hline \hline
\end{tabular} 
\caption{For $\ell=2$ the expected fundamental frequency from linearized theory is $0.37367-0.08896i$, while for $\ell=4$ frequency it is $0.80918-0.09416i$ 
~\cite{Kokkotas99a_url}. Shown are the fitted values from our numerical solutions to the second order Zerilli equations, for four different
sets of initial data. For each initial data configuration we vary the location $r_0$
and width $\sigma$ of the linearized perturbations, keeping one fixed ($r_0=20M$ and $\sigma=4M$, respectively) while varying the other. The fitted frequencies appear to be insensitive to the choice of initial data and agree with the QNM frequencies from linearized perturbation theory.}
\label{tab:bigtable}
\end{table}
\end{center}
\end{widetext}

\bibliography{references}

\end{document}